\documentclass[vecphys]{svmult}

\usepackage{makeidx}         
\usepackage{graphicx}        
\usepackage{multicol}        
\usepackage[bottom]{footmisc}

\makeindex             

\newcommand{\OmegaK}{\Omega_{\rm K}}

\newcommand{\gapprox}{\lower.4ex\hbox{$\;\buildrel >\over{\scriptstyle\sim}\;$}}
\newcommand{\lapprox}{\lower.4ex\hbox{$\;\buildrel <\over{\scriptstyle\sim}\;$}}
\newcommand{\begeq}{\begin{equation}}
\newcommand{\fineq}{\end{equation}}
\newcommand{\msun}{M_\odot} 
 
\newcommand{\betac}{\beta_{\rm crit}}

\def\ellprime0{\ell'_0}

\begin{document}

\title*{Hybrid viscosity and the magnetoviscous instability in hot, collisionless accretion disks}
\titlerunning{Hybrid viscosity and the MVI}
\author{Prasad Subramanian\inst{1}\and
Peter A. Becker\inst{2}\and Menas Kafatos\inst{2}}
\authorrunning{Subramanian, Becker \& Kafatos}
\institute{Indian Institute of Astrophysics, Bangalore - 560034, India
\texttt{psubrama@iiap.res.in}
\and College of Science, George Mason University, Fairfax, VA 22030, USA}
%
%
\maketitle 


\begin{abstract} 
We aim to illustrate the role of hot protons in enhancing the
magnetorotational instability (MRI) via the ``hybrid'' viscosity,
which is due to the redirection of protons interacting with static
magnetic field perturbations, and to establish that it is the only
relevant mechanism in this situation. It has recently been shown by
Balbus \cite{PBM1} and Islam \& Balbus \cite{PBM11} using a fluid
approach that viscous momentum transport is key to the development of
the MRI in accretion disks for a wide range of parameters. However,
their results do not apply in hot, advection-dominated disks, which
are collisionless. We develop a fluid picture using the hybrid
viscosity mechanism, that applies in the collisionless limit. We
demonstrate that viscous effects arising from this mechanism can
significantly enhance the growth of the MRI as long as the plasma
$\beta \gapprox 80$.  Our results facilitate for the first time a
direct comparison between the MHD and quasi-kinetic treatments of the
magnetoviscous instability in hot, collisionless disks.
\end{abstract} 
 
\section{Introduction} 
\label{sec:1} 
The microphysical source of viscosity in accretion disks has been a
long-standing puzzle. Since the early 1990s, there has been a growing
consensus that magnetic fields generated by the magnetorotational
instability (MRI) are key to providing the required viscosity in cold
accretion disks (\cite{PBM2} (and references
therein), \cite{PBM4}, \cite{PBM14}). The standard
treatment of the MRI is valid only for collisional plasmas, which can be
described in the MHD approximation. However, the plasmas comprising hot,
two-temperature accretion flows, like those described in \cite{PBM8} and \cite{PBM19} (hereafter SLE) are clearly collisionless. This is also the case for the radiatively
inefficient, advection-dominated accretion flows (ADAFs) treated in
\cite{PBM15} and \cite{PBM16}.
Filho (\cite{PBM9}), Kafatos (\cite{PBM12}) and Paczy\'nski (\cite{PBM17}) had initially
suggested that viscosity due to collisions between hot protons might be
important in two-temperature accretion flows, although the effects of an
embedded turbulent magnetic field were not included in their treatments.
Subramanian, Becker \& Kafatos (\cite{PBM23}; hereafter SBK96) proposed that a
{\it hybrid} viscosity, due to protons colliding with magnetic
scattering centers, might be the dominant viscosity mechanism in such
accretion disks. In this paper we investigate the implications of the
hybrid viscosity for the development of the MRI in hot disks. In
particular, we show that this mechanism can be used to establish an
interesting connection between the fluid models and the quasi-kinetic
treatments used by previous authors to study the viscous enhancement
of the growth rate during the early stages of the MRI.
\section{MVI in hot accretion disks}

Balbus (\cite{PBM1}) and Islam \& Balbus (\cite{PBM11}) employed an MHD approach to
study the effect of viscosity on the development of the MRI, and
discovered a robust instability which they call the magnetoviscous
instability (MVI). In the MVI, angular momentum is exchanged between
fluid elements via viscous transport, which plays a central role in the
development of the instability. Balbus (\cite{PBM1}) does not address the
physical origin of the viscosity that is central to the development of
the MVI, and therefore his results are stated in terms of an unspecified
coefficient of dynamic viscosity, $\eta$. Islam \& Balbus (\cite{PBM11}) assumed
the plain Spitzer (collisional) viscosity due to proton-proton
collisions in their treatment of the MVI, but this particular mechanism
is not effective in hot, collisionless disks. There have been some
recent attempts at quasi-kinetic treatments of MRI-like instabilities in
collisionless plasmas (e.g., \cite{PBM18}, \cite{PBM20}, \cite{PBM21}). It is interesting to note that the pressure anisotropy concept discussed in these papers is
somewhat similar to the idea embodied in the hybrid viscosity formalism
of SBK96. This suggests that it may be possible to develop a ``fluid''
picture based on the hybrid viscosity that would be applicable in hot
disks, hence bridging the gap between the two paradigms. The hybrid
viscosity concept of SBK96 relies only on the momentum deposited by
particles propagating along magnetic fields lines between adjacent
annuli in the disk. 

\section{Applicability of the hybrid viscosity}

Paczy\'nski (\cite{PBM17}) and SBK96 noted that the presence of even a very weak
magnetic field can effectively ``tie'' protons to magnetic field lines.
Paczy\'nski argued that in this situation the ion-ion collisional mean
free path is much larger than the proton Larmor radius and therefore the
effective mean free path is equal to the proton Larmor radius. This led
him to conclude that the viscosity would effectively be quenched in such
a plasma. However, the protons in hot accretion disks are typically
super-Alfv\'enic, especially in the initial stages of a magnetic
field-amplifying instability such as the MRI, when the plasma $\beta$
parameter is quite large. This reflects the fact that the ratio of the
proton thermal speed to the Alfv\'en speed is equal to $(3\beta/2)^{1/2}$.
Since the magnetic field evolves on Alfv\'en timescales, it can be
considered to be static for our purposes. The motion of collisionless, 
charged particles propagating through a static, tangled magnetic field
has been explored extensively in the context of cosmic ray propagation
(e.g., \cite{PBM5}, \cite{PBM6}, \cite{PBM10}). It has been conclusively established that the
particle transport does not obey Bohm diffusion for a wide range of
rigidities and turbulence levels (see, e.g., Fig.~4 of \cite{PBM5} and Fig.~4 of\cite{PBM6}). In particular,
the low rigidity, low turbulence level case appropriate for our
situation obeys the predictions of quasi-linear theory quite well, and
the mean free paths are much larger than the Larmor radius as expected.

Under these conditions, SBK96 demonstrated the importance of a new kind
of viscosity called the ``hybrid viscosity,'' in which angular momentum
is transported via collisions between protons and static irregularities
(``kinks'') in the magnetic field. In this picture, a proton spirals
tightly along a magnetic field line until its gyro-averaged guiding
center motion (and hence its gyro-averaged momentum) is changed via an
encounter with a kink. During the encounter the proton therefore
exchanges angular momentum with the field, which transfers the resulting
torque to the plasma. The effective mean free path used in the
computation of the viscosity is set equal to the distance between the
magnetic kinks (i.e., the field coherence length). We express the hybrid
viscosity mechanism in terms of a pressure anisotropy in \S~6.1.

Here we examine the implications of the hybrid viscosity for the
development of the MVI in hot, two-temperature accretion disks around
underfed black holes. We assume that the accreting plasma is composed of
fully ionized hydrogen. The physical picture involves the perturbation of
an initially straight magnetic field line that eventually leads to the
instability (see, e.g., Fig.~1 of \cite{PBM1}). Since the proton
Larmor radius is negligible in comparison to a macroscopic length scale,
we can effectively think of the proton as sliding along the field line
like a bead on a wire. The proton is forced to change its direction upon
encountering the kink associated with the initial field perturbation. In
such a situation, the effective mean free path, $\lambda$, used in the
description of the hybrid viscosity should be set equal to the
wavelength of the initial perturbation. We demonstrate that the hybrid
viscosity is the principle mediator of the MVI during the early stages
of the instability.

\section{Hybrid viscosity in hot accretion disks}

The structure of hot, two-temperature accretion disks was first studied
in detail by SLE, and later by Eilek \&
Kafatos (\cite{PBM8}) and SBK96. The closely related advection-dominated
accretion flows were analyzed by Narayan \& Yi (\cite{PBM15}), Narayan,
Mahadevan, \& Quataert (\cite{PBM16}), and many subsequent authors. In this
section, we investigate the nature of the viscosity operative in hot,
two-temperature accretion disks based on a simplified set of
model-independent relations that are applicable to both ADAF and SLE
disks.

The gas in the disk can be considered collisionless with respect to
the protons provided
\begin{equation}
\lambda_{ii} > H \ ,
\label{eq1}
\end{equation}
where $H$ is the half-thickness of the disk, and the ion-ion Coulomb
collisional mean free path, $\lambda_{ii}$, is given in cgs units by
(SBK96)
\begin{equation}
\lambda_{ii} = 1.80 \times 10^5 \, {T_i^{2} \over N_i \, \ln\Lambda}
\ , \label{eq2}
\end{equation}
for a plasma with Coulomb logarithm $\ln\Lambda$ and ion temperature and
number density $T_i$ and $N_i$, respectively. We can combine
equations~(\ref{eq1}) and (\ref{eq2}) to obtain
\begin{equation}
{\lambda_{ii} \over H} = 1.20 \times 10^{-19} \, {T_i^2 \over
\tau_{es} \, \ln\Lambda} \ > \ 1 \ ,
\label{eq3}
\end{equation}
where the electron scattering optical thickness, $\tau_{es}$, is given by
\begin{equation}
\tau_{es} = N_i \, \sigma_{_{\rm T}} \, H \ ,
\label{eq4}
\end{equation}
and $\sigma_{_{\rm T}}$ is the Thomson scattering cross section.
Equation~(\ref{eq3}) can be rearranged to obtain a constraint on
$\tau_{es}$ required for the disk to be collisionless, given by
\begin{equation}
\tau_{es} \ < \ {1.20 \times 10^5 \, T_{12}^2 \over \ln\Lambda}
\ \sim \ 4 \times 10^3 \ ,
\label{eq5}
\end{equation}
where $T_{12}\equiv T/10^{12}\,$K and the final result holds for
$\ln\,\Lambda=29$ and $T_{12}\sim 1$. This confirms that tenuous,
two-temperature disks with $T_i \sim 10^{11}$--$10^{12}\,$K will be
collisionless for typical values of $\tau_{es}$.

The collisionless nature of hot two-temperature accretion flows
established by equation~(\ref{eq5}) strongly suggests that the plain
Spitzer viscosity is not going to be relevant for such disks, although
the answer will depend on the strength of the magnetic field. The hybrid
viscosity will dominate over the Spitzer viscosity provided the ion-ion
collisional mean free path $\lambda_{ii}$ exceeds the Larmor radius,
$\lambda_{\rm L}$, so that the protons are effectively ``tied'' to
magnetic field lines. We therefore have
\begin{equation}
\lambda_{ii} > \lambda_{\rm L} \ ,
\label{eq6}
\end{equation}
where the Larmor radius is given in cgs units by (SBK96)
\begin{equation}
\lambda_{\rm L} = 0.95 \, {T_i^{1/2} \over B} \ ,
\label{eq7}
\end{equation}
where $B$ is the magnetic field strength. Whether the disk is of the SLE
or ADAF types, it is expected to be in vertical hydrostatic equilibrium,
and therefore
\begin{equation}
H \OmegaK = c_s = \sqrt{k T_i \over m_p} \ ,
\label{eq8}
\end{equation}
where $\OmegaK=(GM/r^3)^{1/2}$ is the Keplerian angular velocity
at radius $r$ around a black hole of mass $M$, $c_s$ is the
isothermal sound speed, and $k$ and $m_p$ denote Boltzmann's constant
and the proton mass, respectively.

We can utilize equation~(\ref{eq6}) to derive a corresponding constraint
on the plasma $\beta$ parameter,
\begin{equation}
\beta \equiv {8 \pi N_i k T_i \over B^2} \ ,
\label{eq9}
\end{equation}
such that the hybrid viscosity dominates over the Spitzer viscosity. By
combining equations~(\ref{eq2}), (\ref{eq4}), (\ref{eq6}), (\ref{eq7}),
(\ref{eq8}), and (\ref{eq9}), we find that
\begin{equation}
\beta \ < \ 3.71 \times 10^{32} \, {T_{12}^{9/2} \, R^{3/2} \, M_8
\over \tau_{es} \, (\ln\Lambda)^2} \ ,
\label{eq10}
\end{equation}
where $M_8 \equiv M/(10^8 \msun)$ and $R\equiv r c^2/(GM)$. The minimum
possible value of the right-hand side in equation~(\ref{eq10}) is
obtained for the maximum value of $\tau_{es}$, which is given by
equation~(\ref{eq5}). We therefore find that
\begin{equation}
\beta \ < \ 3.09 \times 10^{27} \, {T_{12}^{5/2} \, R^{3/2} \, M_8
\over \ln\Lambda} \ .
\label{eq11}
\end{equation}
This relation is certainly satisfied in all cases involving the
accretion of plasma onto a black hole, even in the presence of an
infinitesimal magnetic field. We therefore conclude that the protons
will be effectively tied to the magnetic field lines in two-temperature
accretion disks around stellar mass and supermassive black holes, which
implies that the hybrid viscosity dominates over the Spitzer viscosity
in either SLE or ADAF disks.

The results of this section confirm that protons in two-temperature
accretion disks rarely collide with each other, and are closely tied to
magnetic field lines, even for very weak magnetic fields. If a field
line is perturbed, a typical proton sliding along it will follow the
perturbation, and will thus be effectively redirected. This is the basic
premise of the hybrid viscosity concept, which we will now apply to the
development of the MVI.

\section{MVI driven by the hybrid viscosity}

Figure~1 of \cite{PBM11} shows that magnetoviscous effects
significantly enhance the MRI growth rates in the parameter regime
\begin{equation}
X \lapprox x \ , \ \ \ \ Y \gapprox y \ ,
\label{eq12}
\end{equation}
where $x \sim 1$, $y \sim 1$, and 
\begin{equation}
\nonumber
X \equiv {2.0 \, (k_Z \, H)^2 \over \beta} \ , \ \ \ \
Y \equiv {1.5 \, \eta \, k_\perp^2 \over N_i \, m_p \, \OmegaK} \ ,
\label{eq13}
\end{equation}
with $\eta$ denoting the coefficient of dynamic viscosity and $k_Z$ and
$k_\perp$ representing the $z$ and transverse components of the field
perturbation wavenumber, respectively. The maximum MVI growth rate is
$\sqrt{3}\,\OmegaK$, which is $4/\sqrt{3} \sim 2.3$ times larger than
the maximum MRI growth rate of $(3/4)\,\OmegaK$. The conditions in
equation~(\ref{eq12}) are derived from the dispersion relation given in
equation~(33) of \cite{PBM11}, which is general enough to
accommodate different prescriptions for the viscosity coefficient
$\eta$. The condition $X \lapprox x$ implies a constraint on $\beta$
given by
\begin{equation}
\beta \gapprox {2 \, (k_Z \, H)^2 \over x} \ .
\label{eq14}
\end{equation}

As mentioned earlier, a proton sliding along a given field line is
forced to change its direction when it encounters a kink/perturbation in
the field line. The effective viscosity arises due to the momentum
deposited in the fluid by the proton when it encounters the
perturbation. In this picture, the perturbation wavelength plays the
role of an effective mean free path. If we consider perturbations along
an initially straight field line, as in Figure~1 of \cite{PBM1},
then only the transverse component of the perturbation wavelength is
relevant, and the effective mean free path for the proton is therefore
\begin{equation}
\lambda = {2 \, \pi \over k_\perp} \equiv \xi H \ ,
\label{eq15}
\end{equation}
where $\xi \le 1$, since the perturbation wavelength $\lambda$ cannot exceed
the disk half-thickness $H$ (SBK96).

In general, the Shakura-Sunyaev (\cite{PBM22}) viscosity parameter $\alpha$ is
related to the coefficient of dynamic viscosity $\eta$ via (SBK96)
\begin{equation}
\alpha P \equiv - \eta \, R \, {d \OmegaK \over d R}
= {3 \over 2} \, \eta \, \OmegaK \ ,
\label{eq16}
\end{equation}
where $P=N_i \, k \, T_i$ is the pressure in a two-temperature disk with
$T_i \gg T_e$. By combining equations~(\ref{eq8}), (\ref{eq13}), and
(\ref{eq16}), we find that the condition $Y \gapprox y$ can be rewritten
as
\begin{equation}
(k_\perp \, H)^2 \, \alpha \gapprox y \ .
\label{eq17}
\end{equation}
Following Islam \& Balbus (\cite{PBM11}), we expect that $k_\perp \lapprox k_Z$.
By combining equations~(\ref{eq14}) and (\ref{eq17}), we therefore
conclude that $\beta$ must satisfy the condition
\begin{equation}
\beta \gapprox {2 \, y \over \alpha x} \ .
\label{eq18}
\end{equation}
We can also combine equations~(\ref{eq14}) and (\ref{eq15}) to obtain
the separate constraint
\begin{equation}
\beta \gapprox {79 \over \xi^2 x} \ .
\label{eq19}
\end{equation}
Equations~(\ref{eq18}) and (\ref{eq19}) must {\it both} be satisfied if
the MVI is to significantly enhance the MRI growth rates. Hence the
combined condition for $\beta$ is given by
\begin{equation}
\beta \gapprox {\rm Max}\left({79 \over x \xi^2} \ , \ {2 \, y \over \alpha x}
\right) \ .
\label{eq20}
\end{equation}

We can use equation~(\ref{eq16}) to calculate the Shakura-Sunyaev
parameter $\alpha_{\rm hyb}$ describing the hybrid viscosity. The
associated coefficient of dynamic viscosity is given by 
\begin{equation}
\eta_{\rm hyb} = {\lambda \over \lambda_{ii}}
\, \eta_{_{\rm S}} \ ,
\label{eq21}
\end{equation}
where $\lambda_{ii}$ is computed using equation~(\ref{eq2}) and
$\eta_{_{\rm S}}$ is the standard Spitzer collisional viscosity,
evaluated in cgs units using
\begin{equation}
\eta_{_{\rm S}} = 2.20 \times 10^{-15} \, {T_i^{5/2} \over \ln\Lambda} \ .
\label{eq22}
\end{equation}
The quantity $\eta_{\rm hyb}$ defined in equation~(\ref{eq21}) describes
the effect of momentum deposition due to protons spiraling tightly along
a magnetic field line over a mean free path $\lambda$. It differs from
the expression given in equation~(2.14) of SBK96 by a factor of $2/15$,
because we do not consider tangled magnetic fields here. Setting $\eta =
\eta_{\rm hyb}$ in equation~(\ref{eq16}) and utilizing
equations~(\ref{eq2}), (\ref{eq8}), (\ref{eq15}), (\ref{eq21}), and
(\ref{eq22}), we find after some algebra that the expression for
$\alpha_{\rm hyb}$ reduces to the simple form
\begin{equation}
\alpha_{\rm hyb} = 1.2 \, \xi
\ .
\label{eq23}
\end{equation}
We can now combine equations~(\ref{eq20}) and (\ref{eq23}) to conclude
that in the case of the hybrid viscosity, the MVI is able to effectively
enhance the MRI growth rates if
\begin{equation}
\beta \gapprox \betac \equiv {\rm Max}\left({79 \over x \xi^2} \ ,
\ {1.7 y \over x \xi}\right)
\ .
\label{eq24}
\end{equation}
In particular, we note that if $x \sim 1$ and $y \sim 1$, then
equation~(\ref{eq24}) reduces to $\betac = 79\,\xi^{-2}$, since
$\xi \le 1$. We therefore conclude that magnetoviscous effects driven by
the hybrid viscosity will significantly enhance the growth rate
(compared with the standard MRI growth rate) until the plasma $\beta$
parameter reaches $\sim 80$, or, equivalently, until the field strength
$B$ reaches $\sim 10\%$ of the equipartition value. This assumes that
the dominant perturbations have $\xi \sim 1$, which is expected to be
the case during the early stages of the instability. Once the field
exceeds this strength, the growth rate of the instability during the
linear stage will be equal to the MRI rate.

\section{Relation to previous work}

It is interesting to contrast our result for the $\beta$ constraint with
those developed by previous authors using different theoretical
frameworks.

\subsection{Hybrid viscosity in terms of pressure anisotropy}

Before proceeding on to discussing the result for the $\beta$
constraint, we first cast the basic hybrid viscosity mechanism in terms
of a pressure anisotropy. Several similar treatments appeal to a
large-scale pressure anisotropy, rather than an explicit viscosity
mechanism (e.g., \cite{PBM18}, \cite{PBM20}, \cite{PBM21}). It is therefore instructive to show
that the hybrid viscosity mechanism we employ can be cast in these
terms.

We follow the approach of SBK96 in considering a perturbation in the
local magnetic field of an accretion disk. The pressure anisotropy due
to the momentum flux carried by the particles can be analyzed in the
local region using cartesian coordinates, with the $\hat{z}$-axis
aligned in the azimuthal (orbital) direction, the $\hat{y}$-axis
pointing in the outward radial direction, and the $\hat{x}$-axis
oriented in the vertical direction. The unperturbed magnetic field is
assumed to lie in the $\hat{z}$ direction, and the perturbed field makes
an angle $\theta$ with respect to the $\hat{z}$-axis, and an azimuthal
angle $\phi$ with respect to the $\hat{x}$-axis. In keeping with the
hybrid viscosity scenario, we assume that the particles spiral tightly
around the perturbed field line. In this situation, the component of the
particle pressure in the direction {\it parallel} to the magnetic field,
$P_{||}$, is equal to the $\hat{z}$-directed flux of the
$\hat{z}$-component of momentum, $P_{zz}$. Likewise, the total particle
pressure {\it perpendicular} to the field, $P_{\perp}$, is equal to the
sum of the $\hat{x}$-directed momentum in the $\hat{x}$-direction and
the $\hat{y}$-directed momentum in the $\hat{y}$-direction, denoted by
$P_{xx}$ and $P_{yy}$, respectively. Following the same approach that
leads to equation~(2.11) of SBK96, we obtain for the parallel pressure
\begin{eqnarray}
\nonumber
P_{||} = P_{zz} = 2\,m_p\,N_{i}\,{\rm cos}^{2} \theta \, \times \\
\biggl [\frac{k T_{i}}{2 m_p}
- \biggl ( \frac{2 k T_{i}}{\pi m_p} \biggr )^{1/2} \, u'(0) \, \lambda \,
\cos\theta \, \sin\theta \, \sin\phi \biggr ]
\ ,
\label{eq25}
\end{eqnarray}
where $u(y)$ represents the shear velocity profile and the prime denotes
differentiation with respect to $y$.
Similarly, the total perpendicular pressure is given by
\begin{eqnarray}
\nonumber
P_{\perp} = P_{xx} + P_{yy} = 2\,m_p\,N_{i}\, \sin^2\theta \, \times \\
\biggl [\frac{k T_{i}}{2 m_p}
- \biggl ( \frac{2 k T_{i}}{\pi m_p} \biggr )^{1/2} \, u'(0) \, \lambda \,
\cos\theta \, \sin\theta \, \sin\phi \biggr ] \ .
\label{eq26}
\end{eqnarray}
Taken together, equations~(\ref{eq25}) and (\ref{eq26}) imply that
\begin{equation}
\frac{P_{\perp}}{P_{||}} = \tan^2 \theta \ .
\label{eq27}
\end{equation}
This result characterizes the pressure anisotropy associated with the
hybrid viscosity mechanism. Equation~(\ref{eq27}) is strictly valid only
in the limit of zero proton gyroradius, which is a reasonable
approximation in hot advection-dominated disks. When the field line is
unperturbed, so that it lies precisely along the $\hat{z}$-direction,
then $\theta=0$, and equation~(\ref{eq27}) indicates that the
perpendicular pressure tends to zero; in reality, owing to finite
gyroradius effects, the perpendicular pressure would actually be a
small, but finite quantity even in this limit. Early in the instability,
when the field line is slightly perturbed, $\theta$ has a small but
non-zero value, and equation~(\ref{eq27}) predicts that the
perpendicular pressure starts to increase in relation to the parallel
pressure. We have cast the hybrid viscosity mechanism in terms of a
pressure anisotropy in this section in order to make contact with that
part of the literature in which viscous momentum transport is treated
solely in this manner. The quasi-kinetic treatments of Quataert and
co-workers rely on a Landau fluid closure scheme for deriving the
perturbed pressure. The pressure anisotropy implied by the hybrid
viscosity mechanism (eq.~[\ref{eq27}]) is much simpler than the
corresponding result obtained using either the fluid closure scheme of
Quataert et al., or the double adiabatic scheme (\cite{PBM7}) 
adopted by other authors.

\subsection{Relation to MVI treatment}

In their treatment of the MVI, Islam \& Balbus (\cite{PBM11}) parametrized the
viscous transport in terms of an unspecified proton-proton collision
frequency, $\nu$. Their estimates of the growth rates in {\it
collisional} plasmas agree fairly well with those derived using
quasi-kinetic treatments. Based on their formalism, they conclude that
the $\beta$ regime within which magnetoviscous effects can significantly
impact the MRI growth rates in two-temperature accretion flows extends
to $\betac \sim 1$. However, as they point out, their approach breaks
down in the collisionless limit $\nu \to 0$, which describes the ADAF
disks of interest here. It is therefore not surprising that their
constraint on $\beta$ is significantly different from the one we have
derived in equation~(\ref{eq24}).

\subsection{Relation to quasi-kinetic treatment}

Quataert, Dorland \& Hammett (\cite{PBM18}) have treated the case of a strictly
collisionless plasma using a fairly complex kinetic formalism. Their
results suggest that, for the case with $B_{\phi} = B_{z}$ and $k_{r} =
0$ (which is the one considered by Islam \& Balbus and ourselves),
viscous effects will significantly impact the MRI growth rates for
values of $\beta$ that are several orders of magnitude larger than those
predicted by our formalism. For example, their analysis predicts that a
growth rate of $1.5\, \OmegaK$ can be achieved if $\beta \gapprox \betac
\sim 10^4$ (see Fig.~4 of \cite{PBM18}) and Fig.~2 of \cite{PBM20}). On the other hand, Figure~1 of
\cite{PBM11} indicates that a growth rate of $1.5\, \OmegaK$
can be achieved in the MHD model if $X \lapprox 0.35$, $Y \gapprox 12$,
which corresponds to $x = 0.35$, $y = 12$ in equation~(\ref{eq12}).
Assuming that $\xi \sim 1$ as before, equation~(\ref{eq24}) yields in
this case the condition $\beta \gapprox \betac \sim 225$. Hence our MHD
model based on the hybrid viscosity predicts that viscous effects will
enhance the MRI growth rates down to much lower values of $\beta$ than
those obtained in the quasi-kinetic model. This difference reflects the
differing role of the particle pressure in the two scenarios.

In our formulation, the viscosity is expressed by protons that deposit
their momentum into the fluid upon encountering kinks in the magnetic
field, which is anchored in the local gas. The importance of forces due
to gas pressure relative to those due to the tension associated with the
magnetic field thus scales as the plasma $\beta$. On the other hand, gas
pressure forces are only $\sqrt{\beta}$ times as important as forces
arising out of magnetic tension in the quasi-kinetic treatment of
\cite{PBM18}. It follows that the value of
$\betac$ computed using our MHD model based on the hybrid viscosity
should be comparable to the square root of the $\betac$ value obtained
using the quasi-kinetic model, and this is borne out by the numerical
results cited above.

\section{Conclusions}

In this paper we have investigated the role of hot protons in
influencing the magnetoviscous instability described in \cite{PBM1}
and \cite{PBM11}. We have shown that the only relevant
viscosity mechanism in this situation is the ``hybrid'' viscosity, which
is due to the redirection of protons interacting with magnetic
irregularities (``kinks'') set up by the initial field perturbations. In
particular, we have demonstrated in equation~(\ref{eq24}) that viscous
effects associated with the hybrid viscosity will significantly augment
the MRI growth rates for $\beta \gapprox 80$, which corresponds to a
magnetic field strength $B$ below $\sim 10\%$ of the equipartition
value. For smaller values of $\beta$, we expect the instability to grow
at the MRI rate as long as it remains in the linear regime. This
conclusion is expected to be valid in any hot, two-temperature accretion
disk, including advection-dominated ones. We have obtained this result
using a relatively simple fluid treatment, based upon the general
dispersion relation obtained in \cite{PBM11}. Our use of the
hybrid viscosity concept alleviates an important drawback in the fluid
application made by Islam \& Balbus (\cite{PBM11}), because their treatment of
viscous transport breaks down in the collisionless plasmas of interest
here. The new results we have obtained allow an interesting comparison
between the MHD approach and the quasi-kinetic formalism used by other
authors. We show that the differences between the predictions made by
the two methodologies stem from the differing treatments of the particle
pressure.

PS gratefully acknowledges the hospitality of the Jagannath Institute of
Technology and Management, where part of this work was carried out.

%
%
%
%
%

%
%



\printindex
\end{document}